\documentclass[aps,prb,twocolumn,showpacs,twoside,floatfix,superscriptaddress]{revtex4}
\usepackage{graphicx}
\usepackage{graphics}
\usepackage{amsmath}
\usepackage{amssymb}
\usepackage{mathrsfs}
\usepackage{epsf}

\begin{document}

\title{Simple model for transport phenomena : Microscopic construction of
Maxwell Demon like engine}

\author{Jyotipratim Ray Chaudhuri}
\email{jprc_8@yahoo.com}
\affiliation{Department of Physics, Katwa College,
Katwa, Burdwan 713130, India}

\author{Sudip Chattopadhyay}
\email{sudip_chattopadhyay@rediffmail.com}
\affiliation{Department of Chemistry, Bengal Engineering and
Science University, Shibpur, Howrah 711103, India}

\author{Suman Kumar Banik}
\email{skbanik@phys.vt.edu}
\affiliation{Department of Physics,
Virginia Polytechnic Institute and State University,
Blacksburg, VA 24061-0435, USA}

\date{November 08, 2007}

\begin{abstract}
We present a microscopic Hamiltonian framework to develop Maxwell
demon like engine. Our model consists of a equilibrium thermal bath
and a non-equilibrium bath; latter generated by driving with an
external stationary, Gaussian noise. The engine we develop, can be
considered as a device to extract work by modifying internal
fluctuations. Our theoretical analysis focusses on finding the
essential ingredients necessary for generating fluctuation induced
transport under non-equilibrium condition. An important outcome of
our model is that the net motion occurs when the non-linear bath is
modulated by the external noise, creating the non-zero effective
temperature even when the temperature of both the baths are same.
\end{abstract}


\maketitle

\section{Introduction}
It is well known from elementary text books of thermodynamics that
it is possible to extract some amount of mechanical work from a
thermal bath at a temperature $\overline{T}$ provided there is
another bath at a lower temperature $T$ ($\overline{T} > T$).
Thermal engines are the devices that perform this task. We also know
that any object in a thermal bath follows principle of equipartition
of energy and exhibits random energy fluctuations of the order of
$k_BT$. In the macroscopic scale this fluctuations are very small
but of very import relevance for nanoscopic objects such as
biological motors \cite{oster}. The question one can raise, if it is
possible to rectify thermal fluctuations by some appropriate
mechanical devices, e.g., Maxwell's demon like engine or Carnot
engine. Maxwell's demon manages to decrease the entropy, in other
words, it increases the amount of energy available by increasing its
knowledge about the motion of all the molecules. Thermodynamics says
this is impossible, one can only increase entropy. To resolve this
paradox (violation of second law of thermodynamics), an inchoate
relationship between information and energy emerged. This problem
has led to very interesting links between physics, information
theory and the theory of computation - from concepts of information
entropy to reversible computing. Feynman used a ratchet example to
illustrate some implications of the second law of thermodynamics
\cite{feynmann} and now it is quite well known that useful work
cannot be extracted from systems undergoing equilibrium
fluctuations. Under thermal equilibrium situation, no net particle
current can be generated in the presence of external potential of
arbitrary shape in accordance with the principle of detail balance.
On the other hand, in non-equilibrium situation, due to the
breakdown of the principle of detail balance net current flow is
possible and thus one can transform energy into work at the expense
of increased entropy \cite{reimann}. The study of such a process
from a microscopic point of view is quite challenging and hence an
important topic in numerous areas of molecular sciences because no
principles of generality that hold for equilibrium cases are
applicable in non-equilibrium situations \cite{astumian,magnasco}.
To explain such processes from theoretical point of view, various
types of models have been proposed in the literature
\cite{feynmann,reimann,astumian,magnasco}. Though most of them are
phenomenological, models from first principle have also been
proposed in this direction \cite{millonas,jayannavar}.

Brownian motion cannot create a steady flux in a system in
equilibrium. Nor can local asymmetries in a static potential energy
landscape (simple example of non-equilibrium situation) rectify
Brownian motion to induce a drift. A landscape that varies in time,
however, can eke a flux out of random fluctuations by breaking
spatiotemporal symmetry. Such flux-inducing time dependent
potentials are known as thermal ratchets
\cite{reimann,magnasco,thermalratchet}, and their ability to bias
diffusion by rectifying thermal fluctuations has been proposed as a
possible mechanism for transport by molecular motors and is being
actively exploited for macromolecular separation
\cite{motor,julicher}.

Most thermal ratchet models are based on spatially asymmetric
potentials. Their time variation involves displacing or tilting them
relative to the laboratory frame, modulating their amplitude,
changing their periodicity, or some combination, usually in a
two-state cycle. It is now well known that a spatially symmetric
potential still can induce drift in a cycle of three states, one of
which allows for free diffusion. More recently, directed transport
has been induced in an atomic cloud by a spatially symmetric rocking
ratchet created with an optical lattice \cite{ref12}.

In this paper we investigate a self-consistent fluctuation induced
transport theory through a microscopically constructed Maxwell demon
type information engine. In our theory, the system under
consideration is coupled to two independent baths maintained at two
different equilibrium temperatures. A thermal fluctuation in one
bath is created by modulating it externally via a random force,
there by making the whole system thermodynamically open under the
exposure of non-equilibrium fluctuations. We then derive a
mathematical expression for the fluctuation induced transport
current under non-equilibrium situation analytically, which holds
for all temperatures and then apply it to various cases of physical
relevance, particularly, in the calculation of escape rate.

The prime advantage of adopting the external noise driven
non-equilibrium bath in the present formalism is as follows. The
external noise drives the heat bath out of equilibrium; as a result,
a shift in the equilibrium temperature takes place through the
creation of an effective temperature which the system of interest
experiences in the steady state \cite{jrc1,cbb}. Creation of
non-equilibrium state (and of an effective temperature) through
external driving is one of the essential requirements to break the
symmetry of the system of interest that may lead to the generation
of noise induced transport. In support of our definition of
effective temperature, Popov and Hernandez \cite{popov} recently
provided an extensive and elegant analysis for defining generalized
temperature in the context of non-equilibrium open systems.

A number of different situations depicting the modulation of one
bath out of two, may be of physically relevant. As for example, we
may think about heat transfer through a metallic rod when two ends
of the rod are immersed in two (different) liquids kept at different
equilibrium temperatures. The liquids may act as Brownian bath and
the one which is photochemically active, may be exposed to an
external fluctuating light intensity. Since the fluctuations in the
light intensity result in the fluctuations in the polarization of
the liquid molecules, the effective temperature field around the end
of the metal bar gets modified, by making the liquids and rod system
thermodynamically open and by throwing it in a non-equilibrium
situation.

The organization of the paper is the following. In Sec.II we
describe the construction and essential features of the microscopic,
Hamiltonian based model. Analysis for external noise induced
transport has been described in Sec.III. Sec.IV provides a general
analysis for noise induced transport. The paper is concluded in
Sec.V.

\section{The Model}

In our model the system is coupled with two baths $A\{ q_j, p_j\}$ and
$B\{Q_j, P_j\}$ with characteristic frequencies $\{\omega_j\}$ and
$\{\Omega_j\}$, respectively. The coupling between the system and the bath
$A$ and bath $B$ is linear and non-linear in nature, respectively.
The Hamiltonian for the composite system can be written as \cite{zwanzig}
\begin{eqnarray}\label{eq1}
H & = & H_S + H_A + H_{SA} + H_B + H_{SB} + H_{int} \nonumber \\
  & = & \frac{p^2}{2M} + U(x) + \frac{1}{2} \sum_{i=1}^{N} \left \{
\frac{p_i^2}{m_i} +m_i \omega_i^2 (q_i-g_ix)^2 \right \}  \nonumber \\
& & + \frac{1}{2} \sum_{j=1}^{N} \left \{ \frac{P_j^2}{M_j} +M_j \Omega_j^2
(Q_j-c_j f(x) )^2 \right \} +H_{int}
\end{eqnarray}

\noindent with $H_{\rm int} = \sum_{j=1}^{N} \kappa_j Q_j
\epsilon(t)$. The first two terms on the right hand side of Eq.
(\ref{eq1}) represent the system mode and the third term describes
the Hamiltonian of the bath $A$, maintaining the thermal environment
of the engine, and the system bath interaction $(H_A + H_{SA})$. The
fourth term corresponds to the bath $B$ to which the system is
nonlinearly coupled. The bath $B$ is modulated by an external noise
$\epsilon(t)$. $H_{int}$ represents the interaction between the
nonlinear modes $Q_j$ and the external noise $\epsilon(t)$. In
Eq.(\ref{eq1}) $f(x)$ is some smooth function of the system variable
$x$ while $g_i$, $c_j$ and $\kappa_j$ are the coupling constants
between the system and bath $A$, the system and bath $B$ and the
bath $B$ and the external driving force $\epsilon (t)$,
respectively. Both the baths are in thermal equilibrium with
characteristic temperature $T_a$ (for bath $A$) and $T_b$ (for bath
$B$), respectively, in presence of the system. In comparison to the
earlier developments \cite{millonas,jayannavar}, in our model, the
bath $B$ is externally driven by an stationary gaussian noise
agency. The external noise $\epsilon(t)$ has zero mean and arbitrary
decaying correlation function:
\begin{equation}\label{eq2}
\langle \epsilon(t) \rangle = 0 \text{, }
\langle \epsilon(t)\epsilon(t^{\prime})\rangle = 2 D \psi(t-t^{\prime})
\end{equation}

\noindent where $D$ is the external noise strength and $\psi(t)$ is
the memory kernel of external noise $\epsilon(t)$. The physical
situation we address here is that at $t = 0$, both the baths $A$ and
$B$ are in thermal equilibrium in the presence of the system but in
the absence of the external noise agency. At $t = 0_+$, the external
noise agency is switched on and the $B$-bath is modulated by
$\epsilon(t)$.

After eliminating the bath variables, the equations of motion
describing the system dynamics (considering the masses of the system
and reservoir mode to be unity) is written as
\begin{eqnarray}\label{eq3}
\dot{x} &=& v \nonumber \\
\dot{v} &=& - \frac{dU}{dx} - \int_0^t
dt^{\prime} \gamma_a (t-t^{\prime}) v(t^{\prime}) + \xi_a(t)
\nonumber \\
&& -\frac{df}{dx} \int_0^t dt^{\prime} \gamma_b
(t-t^{\prime}) \frac{df(x(t^{\prime}))}{dx(t^{\prime})}
v(t^{\prime}) + \frac{df}{dx} \{\xi_b(t)+\pi(t)\} \nonumber \\
\end{eqnarray}

\noindent where
\begin{eqnarray} \label{eq4}
\gamma_a(t) &=& \sum_{i=1}^N g_i^2 \omega_i^2 \cos{\omega_i t}
\text{, } \gamma_b(t) = \sum_{j=1}^N c_j^2 \Omega_j^2
\cos{\Omega_j t} \nonumber \\
\pi(t) &=& - \int_0^t dt^{\prime}
\varphi(t-t^{\prime}) \epsilon(t^{\prime}) \text{, }
\varphi(t) = \sum_j c_j \Omega_j \kappa_j \sin(\Omega_j t)
\nonumber \\
\xi_a(t) & =&  \sum_{i=1}^N g_i \{ [q_i(0) - g_i x(0)] \omega_i^2
\cos{\omega_i t} + v_i (0) \omega_i \sin(\omega_i t)\}
\nonumber \\
\xi_b(t) &=& \sum_{j=1}^N c_j \{ [Q_j(0) - c_j f(x(0))] \Omega_j^2
\cos{\Omega_j t} \nonumber \\
&& + V_j(0) \Omega_j \sin(\Omega_j t)\}
\end{eqnarray}

\noindent Here $\gamma_a$ and $\gamma_b$ are friction coefficient,
generated due to coupling of the system with the two baths. $\{
q_i(0), v_i(0)\}$ and $\{Q_j(0), V_j(0)\}$ are the initial values of
the two bath variables. $\xi_a(t)$ and $\xi_b(t)$ are two noises due
to the presence of the two baths $A$ and $B$, respectively. The
statistical properties of $\xi_a(t)$ and $\xi_b(t)$ are found to be
\begin{eqnarray}
&& \langle \xi_a(t)\rangle_a = 0 \text{ , } \langle
\xi_a(t)\xi_a(t^{\prime})\rangle_a = k_B T_a
\gamma_a(t-t^{\prime}) \nonumber \\
&& \langle \xi_b(t) \rangle_b = 0 \text{ , }
\langle \xi_b(t)\xi_b(t^{\prime})\rangle_b = k_B T_b
\gamma_b(t-t^{\prime})
\end{eqnarray}\label{eq5}

\noindent where $\langle \cdots\rangle_a$ and $\langle
\cdots\rangle_b$ are ensemble averages over the distributions of
initial bath variables (for baths $A$ and $B$, respectively) which
are assumed to be canonically distributed with distribution
functions:
\begin{subequations}
\begin{eqnarray}
\label{eq6a} \text{Bath A}: && P_a =\frac{1}{Z_a} \exp \left (-
\frac{H_A^0+H_{SA}^0}{k_B T_a} \right )
\\
\label{eq6b} \text{Bath B}: && P_b =\frac{1}{Z_b} \exp \left (-
\frac{H_B^0+H_{SB}^0}{ k_B T_b} \right )
\end{eqnarray}
\end{subequations}

\noindent with $Z_a$ and $Z_b$ being the two normalization constants
and the superscript `0' in the Hamiltonian signifies the bath
coordinates at time $t=0$.

Let us now define an effective noise:
\begin{equation}
\eta (t) = \xi_b(t) + \pi(t).
\end{equation}\label{eq7}

\noindent As $\xi_b(t)$ and $\pi(t)$ are both stationary and
Gaussian, the effective noise $\eta (t)$ will also be stationary
and Gaussian. The statistical properties of $\eta (t)$ is given by
\begin{eqnarray}\label{eq7a}
\langle \langle \eta(t) \rangle \rangle_b &=&0
\nonumber \\
\langle \langle \eta(t)\eta(t^{\prime})\rangle \rangle_b &=& k_B T_a
\gamma_a(t-t^{\prime}) + 2D \int_0^t dt^{\prime\prime}
\int_0^{t^\prime} dt^{\prime\prime\prime}
\nonumber \\
&& \times \varphi(t-t^{\prime\prime})
\varphi(t^{\prime}-t^{\prime\prime\prime})
\psi(t^{\prime\prime}-t^{\prime\prime\prime})
\end{eqnarray}

\noindent where $\langle\langle\cdots\rangle\rangle_b$ means we have
taken two averages, averages of bath ($B$) variables and averages
over each realization of $\epsilon(t)$, independently. It is
important to mention the fact that the last equation, although has a
close kinship with the famous fluctuation-dissipation relation, it
is not the fluctuation-dissipation relation due to the presence of
the external noise $\epsilon(t)$ through $\varphi(t)$. Rather, it
serves as a thermodynamic consistency condition \cite{cbb}.

To obtain finite result in the continuum limit $(N\rightarrow
\infty)$, we replace the summation by integration and consider
density of modes ${\cal D}_a(\omega)$ and ${\cal D}_b(\Omega)$ for
two baths $A$ and $B$, respectively and assume the coupling
functions as \cite{cbb}:
\begin{equation}\label{eq8}
g(\omega) = \frac{g_0}{\omega \sqrt{\tau_a}},
c(\Omega) = \frac{c_0}{\Omega \sqrt{\tau_b}}
\text{ and }
\kappa ({\Omega}) = \kappa_0  \Omega \sqrt{\tau_b}
\end{equation}

\noindent where $g_0$, $c_0$ and $\kappa_0$ are some constants. With
these choices, $\gamma_a(t)$, $\gamma_b(t)$ and $\varphi(t)$ reduces
to
\begin{subequations}
\begin{eqnarray}
\label{eq9a} \gamma_a(t) & = & \frac{g_0^2}{\tau_a} \int d \omega
{\cal D}_a
(\omega) \cos(\omega t), \\
\label{eq9b} \gamma_b(t) & = & \frac{c_0^2}{\tau_b} \int d \Omega
{\cal D}_b
(\Omega) \cos\Omega t, \\
\label{eq9c} \varphi(t) & = & c_0 \kappa_o \int d \Omega {\cal
D}_b(\Omega) \Omega \sin (\Omega t).
\end{eqnarray}
\end{subequations}

\noindent We choose Lorentzian density of modes ${\cal D}_a(\omega)$
and ${\cal D}_b(\Omega)$:
\begin{equation}
{\cal D}_a(\omega) = \frac{2}{\pi} \frac{\tau_a}{1+\omega^2\tau_a^2}
\text{ and } {\cal D}_b(\Omega) = \frac{2}{\pi}
\frac{\tau_b}{1+\Omega^2\tau_b^2}.
\end{equation}\label{eq10}

\noindent These choices resemble broadly the behavior of the
hydrodynamical modes in a macroscopic system \cite{resibois}. With
these forms of ${\cal D}_a(\omega)$, ${\cal D}_b(\Omega)$,
$g(\omega)$, $c(\omega)$ and $\kappa(\omega)$ we have the
expression for $\gamma_a(t)$, $\gamma_b(t)$ and $\varphi(t)$ as
\begin{subequations}
\begin{eqnarray}
\gamma_a(t) &=& (\gamma_a/\tau_a) \exp\left(-t/\tau_a\right),
\\
\gamma_b(t) &=& (\gamma_b/\tau_b) \exp\left(-t/\tau_b\right),
\\
\varphi(t)  &=& (c_o \kappa_0/\tau_b) \exp\left (-t/\tau_b\right)
\end{eqnarray}
\end{subequations}

\noindent where $\gamma_a=g_0^2$ and $\gamma_b=c_0^2$. When the
correlation times of the two baths, ${\tau_a}$ (for bath $A$) and
${\tau_b}$ (for bath $B$) both tend to zero, one obtains a
$\delta$-correlated noise process. Clearly for Markovian internal
dissipation [$\tau_a\rightarrow 0$, $\tau_b \rightarrow 0$] one has
$\gamma_a(t) = 2 \gamma_a \delta(t)$, $\gamma_b(t) = 2 \gamma_b
\delta(t)$ and $\varphi(t) = 2 c_0 \kappa_0 \delta(t)$.

\section{Generic expression for noise induced transport}

\subsection{Bath modulation by external white noise}
At this point, we consider a specific statistical property of the
external noise $\epsilon(t)$ which is considered to be Gaussian,
stationary and $\delta$-correlated noise with strength $D_0^{1/2}$:
\begin{eqnarray}\label{eq13}
\langle \epsilon(t) \rangle  = 0 \text{, } \langle \epsilon(t)
\epsilon(t^{\prime})\rangle = 2 \textit{D}_0 \delta (t-t^{\prime}).
\end{eqnarray}

\noindent Then the property of the \textit{dressed noise} $\pi(t)$
can be written as
\begin{eqnarray}\label{eq14}
\langle\pi(t) \rangle = 0 \text{, }
\langle\pi(t) \pi(t^{\prime}) \rangle = 2 \textit{D}_0 \gamma_b \kappa_0^2
\delta (t-t^{\prime}).
\end{eqnarray}

\noindent and correspondingly the Langevin equation (\ref{eq3})
becomes
\begin{eqnarray}\label{eq15}
\dot{x} = v \text{, }
\dot{v} = - \frac{dU}{dx} - \Gamma(x) v + \xi_a(t) +
f^{\prime}(x)\eta(t)
\end{eqnarray}

\noindent where
\begin{eqnarray*}
&&\Gamma(x) = \gamma_a + \gamma_b [f^{\prime}(x)]^2, \\
&& \langle \xi_a(t)\rangle = 0 \text{, }
\langle \xi_a(t)\xi_a(t^{\prime}) \rangle = 2 \gamma_a k_B T_a
\delta (t-t^{\prime}), \\
&&\langle \langle \eta(t)  \rangle \rangle_b = 0 \text{, } \langle
\langle \eta(t) \eta(t^{\prime})\rangle \rangle_b =  2 \gamma_b [k_B
T_b + D_0 \kappa_0^2]\delta (t-t^{\prime}).
\end{eqnarray*}

\noindent The Langevin equation (\ref{eq15}) describes a
multiplicative noise process with space dependent dissipation.

For the case of large dissipation, one eliminates the fast variables
adiabatically to get a simpler description of the system which is
valid in much slower time scale. When the Brownian particles move in
a bath with constant large dissipation this adiabatic elimination of
fast variables leads to the correct description of the system.
However, in presence of hydrodynamic interaction, i.e., when the
dissipation is position dependent or equivalently, when the noise is
multiplicative with respect to system variables, the conventional
adiabatic elimination of fast variables does not work. Using the
method proposed by Sancho {\it et al.} \cite{sancho}, which is based
on a systematic expansion of the relevant variables in powers of
$\Gamma^{-1}$ and neglecting terms $\cal{O}$ ($\Gamma^{-1}$), the
Fokker-Planck equation corresponding to the Langevin equation
(\ref{eq15}) in the overdamped limit can be obtained as:
\begin{eqnarray}\label{eq18}
\frac{\partial F}{\partial t} & = & \frac{\partial}{\partial x}
\left \{ \frac{U^{\prime}(x)} {\Gamma(x)} F \right \}
+ \gamma_a T_a \frac{\partial}{\partial x}
\left \{ \frac{1}{\Gamma(x)}
\frac{\partial}{\partial x} \frac{1}{\Gamma(x)} F \right \}
\nonumber \\
&& + \gamma_b {\overline{T}}_b \frac{\partial}{\partial x}
\left \{ \frac{f^{\prime}(x)} {\Gamma(x)} \frac{\partial}{\partial x}
\frac{f^{\prime}(x)}{\Gamma(x)} F \right \} \nonumber \\
&& + \gamma_b {\overline{T}}_b \frac{\partial}{\partial x} \left
\{ \frac{f^{\prime}(x) f^{\prime \prime}(x) } {\Gamma^2(x)}  F
\right \}
\end{eqnarray}
where $F(x,t)$ is the probability distribution function.
\noindent In the ordinary Stratonovich description the Langevin
equation corresponding to the Fokker-Planck equation (\ref{eq15}) is
given by
\begin{eqnarray}\label{eq18a}
\dot{x}= -\frac{U^{\prime}(x)}{\Gamma(x)} - \frac{f^{\prime}(x)f^{\prime\prime(x)}}{\Gamma^2(x)}
+ \frac{1}{\Gamma(x)}\eta_A(t) + \frac{f^{\prime}(x)}{\Gamma(x)}\eta_B(t)
\end{eqnarray}

\noindent Here, $\langle\eta_A\rangle=\langle\eta_B\rangle=0$ and
$\langle\eta_A(t)\eta_A(t^{\prime})\rangle = 2\gamma_aT_a
\delta(t-t^{\prime})$, $\langle\eta_B(t)\eta_B(t^{\prime})\rangle =
2\gamma_b{\overline {T_b}} \delta(t-t^{\prime})$ and $\overline{T}_b
= [ T_b + (D_0 \kappa_0^2)]$ is the {\it effective temperature} of
the bath $\{ Q_j,P_j\}$ and we have set $k_B=1$.

To compute the mean velocity, $\langle \dot{x} \rangle$, required to
study the phenomena of fluctuation induced transport, let $U(x)$ and
$f(x)$ be periodic functions of $x$ and are invariant under the same
transformation $x \rightarrow (x+L)$. Then, following Risken
\cite{risken}, the exact mean velocity $\langle \dot{x} \rangle$ of
the system is given by
\begin{equation}\label{eq19}
\langle \dot{x} \rangle =\frac{1-\exp(-\delta)}{\int_0^L dy
\exp[-\phi(y)] \int_y^{y+L} dx^{\prime} G(x')
\exp[\phi(x^{\prime})]}
\end{equation}

\noindent with $\delta = \phi(x)-\phi(x+L)$ and
$G(x)= \Gamma^2(x)/[ T_a\gamma_a + \overline{T}_b \gamma_b f^2(x)]$.
In the above expression $\phi(x)$ is the effective potential
and is given by
\begin{eqnarray}\label{e20}
\phi(x) & = & \int^x \left \{ \frac{U^{\prime}(x) \Gamma(x)}{T_a \gamma_a +
{\overline T_b} \gamma_b [f ^{\prime}(x)]^2 } \right. \nonumber \\
&& + \left. \frac {{\overline T_b} -T_a}{\Gamma(x)} \frac{
2\gamma_a\gamma_b f ^{\prime}(x) f ^{\prime \prime} (x)} {\gamma_a
T_a + \gamma_b {\overline T_b} [f ^{\prime}(x)]^2 } \right \} dx.
\end{eqnarray}

\noindent From Eq.(\ref{e20}) it is clear that when $U^{\prime}(x)$
= 0 and $f^{\prime}(x)$ as well as $f^{\prime\prime}(x)$ both have
same sign, the direction of transport will depend on the relative
sign of $({\overline T}_b-T_a)$. For ${\overline T}_b> T_a$, the
current will flow in one direction, on the other
hand if ${\overline T}_b< T_a$, the direction is reverse. Also it is
easy to verify from equations (\ref{eq19}) that when ${\overline
T}_b=T_a$ the current vanishes identically as $\delta = 0$ in this
case. The nonvanishing $\delta$ makes $\phi(x)$ asymmetric with an
effective slope which leads to the generation of directed motion.

We are now in a position to discuss some important issues related to
our development which will indicate the relationship of our
formalism with the existing ones in the same direction. It is
important to note that the stationary distribution function
associated with Eq.(\ref{eq18}) is given by $F_S(x) = N
\exp(-\phi(x))$, $N$ being the normalization constant. It reduces to
the correct equilibrium distribution under the situation $T_a =
{\overline T_b}$. There will be no net current if $T_a = {\overline
T_b}$, since in such case $\delta=0$ which makes the numerator in
Eq.(\ref{eq19}) equals to zero. Earlier Millanos \cite{millonas} and
Jayannavar\cite{jayannavar} has shown that when the system is
coupled with two baths having different temperatures, a net current
flows from higher to lower temperature. But in our case, this
situation is quite different. If both the baths are initially kept
at the same temperature $T_a = T_b = T$ (say), there will be current
due to the presence of the term $(D_0 \kappa_0^2)$ in the expression
of effective temperature ${\overline T_b}$ which arises due to the
modulation of nonlinear bath by external noise. If the two baths
consist of the same type of oscillators and kept at the same
equilibrium temperature, we can generate a current by externally
driving the nonlinear baths. In such a case, the effective potential
becomes
\begin{eqnarray}
\phi(x) & = & \int^x \left \{ \frac{U^{\prime}(x)
\Gamma(x)}{T\Gamma(x) +
\gamma_b D_0 \kappa_0^2 [f ^{\prime}(x)]^2 } \right. \nonumber \\
&& \left. + \frac{D_0 \kappa_0^2}{\Gamma(x)} (\frac{
2\gamma_a\gamma_b f ^{\prime}(x) f ^{\prime \prime}(x) } {T\Gamma(x)
+ \gamma_b D_0 \kappa_0^2 [f ^{\prime}(x)]^2}) \right \} dx.
\end{eqnarray}\label{eq21}

From the above expression it is clear that $\phi(x)\ne \phi(x+L)$
and even when the two baths are at the same temperature, there will
be a current from $\{Q, P\}$ to $\{q,p \}$ bath. Let the coupling of
the system with $\{Q, P\}$ be linear: $f(x) = x$ and hence
$f^{\prime \prime}(x) =0 $, then the effective potential $\phi(x)$
satisfies the relation $\phi(2\pi) = \phi(0)$ and consequently, {\it
there will be no net current}. As in the models of
Millonas\cite{millonas} and Jayannavar \cite{jayannavar}, one can
easily demonstrate that there is no net current, if $\gamma_a$ or
$\gamma_b$ are zero, as it should be. Finally, we would like to
point out that the expression of current, Eq.(\ref{eq18}) in our
model reduces to the one obtained by Jayannavar, if there is no
external modulation of the heat bath. At the end of this section we
want to mentioned the fact that, instead of modulating the nonlinear
bath, analogous situation would have been created by driving
externally the linear heat bath and then one can obtain the
corresponding expression for $\langle \dot{x}\rangle$ and $\phi(x)$.

\subsection{Bath modulation by external colored noise}

At this juncture, we consider the case where internal dissipation is
Markovian but the external noise $\epsilon(t)$ is Gaussian,
stationary, and exponentially correlated one with zero mean [i.e.
$\langle \epsilon(t)\rangle=0 $ and $\langle \epsilon(t)
\epsilon(t^{\prime})\rangle=(D_e/\tau_e)
\exp(-|t-t^{\prime}|/\tau_e)$]. Consequently, neglecting the
transient terms, the correlation function of the dressed noise
$\pi(t)$ can be found to be \cite{cbb}: $\langle \pi(t)
\pi(t^{\prime})\rangle=(D_e\gamma_b\kappa_0^2/\tau_e)
\exp(-|t-t^{\prime}|/\tau_e)$. Thus the effective noise $\eta(t)$
will also be exponentially correlated: $\langle \langle \eta(t)
\rangle \rangle_b = 0$ and $\langle \langle \eta(t) \eta(t')\rangle
\rangle_b = (D_R/\tau_R) \exp(-|t-t^{\prime}|/\tau_R)$ where $D_R=
\gamma_b (T_b + D_e \kappa_0^2)$ and $\tau_R = (\gamma_b \kappa_0^2
D_e \tau_e)/D_R$. For $\tau_e\rightarrow 0$ we obtain the previous
case of $\delta$-correlated noise process. Now using van-Kampen's
approach \cite{vankampen}, the Fokker-Planck equation in phase space
corresponding to Langevin equation (\ref{eq3}) becomes
\begin{eqnarray}
\label{neq1}
\frac{\partial F}{\partial t} & = & - \frac{\partial}{\partial x}
(vF) \nonumber \\
&& +\frac{\partial}{\partial v} [\Gamma(x)v + U^{\prime}(x)- 2
f^{\prime}(x) f^{\prime\prime}(x) \tau_R D_R] F \nonumber \\
&& + [f^{\prime}(x)]^2 \tau_R D_R \frac{{\partial}^2F}
{\partial x \partial v} \nonumber \\
&& + \left \{ [f^{\prime}(x)]^2D_R - \Gamma(x) [f^{\prime}(x)]^2
\tau_R D_R \right \} \frac{{\partial}^2F} {\partial v^2}.
\end{eqnarray}

\noindent The above equation is valid for small but finite
correlation time $\tau_e$. The term $[f^{\prime}(x)]^2 \tau_R D_R
(\partial^2F/\partial x \partial v)$ is a small non-Markovian
contribution and for small $\tau_R$, we may neglect this term to get
finally the approximate Fokker-Planck equation in phase space,
\begin{eqnarray}
\frac{\partial F}{\partial t} & = &  - \frac{\partial}{\partial x}(vF)
\nonumber \\
&& +\frac{\partial}{\partial v} [\Gamma(x)v + U^{\prime}(x)- 2
f^{\prime}(x) f^{\prime\prime}(x) \tau_R D_R] F
\nonumber \\
&& + \left \{ [f^{\prime}(x)]^2D_R - \Gamma(x) [f^{\prime}(x)]^2 \tau_R
D_R \right \} \frac{{\partial}^2F} {\partial v^2}
\end{eqnarray}\label{nweq1}

\noindent which can equivalently described by the Langevin
equation in the ordinary Stratonovich sense:
\begin{equation}
\dot{x} = v , \dot{v} = - U^{\prime}(x) - \Gamma(x) v + g(x)\eta_e(t)
\end{equation}\label{neq2}

\noindent where $g(x) = f^{\prime}(x) [1-\Gamma(x)\tau_R]^{1/2}$ and
$\eta_e$ is a Gaussian noise with zero mean and $\langle\eta_e(t)
\eta_e(t^{\prime})\rangle = 2D_R \delta(t-t^{\prime})$. Along the
same line as in the first case of $\delta$-correlated noises, the
expression for the average velocity may be calculated from
Eq.(\ref{eq19}) by replacing the effective potential $\phi(x)$ by
the new effective potential $\psi(x)$ given by
\begin{eqnarray}
\psi(x) & = & \int^x \left \{ \frac{U^{\prime}(x) \Gamma(x)}{T_a
\gamma_a +
{\overline T_b} \gamma_b g ^2(x) } \right. \nonumber \\
&& + \left. \frac {{\overline T_b} -T_a}{\Gamma(x)} \left (\frac{
2\gamma_a\gamma_b g(x) g^{\prime} (x)} {\gamma_a T_a + \gamma_b
{\overline T_b} g ^2(x) }\right ) \right \} dx.
\end{eqnarray}\label{neq3}

\noindent When both the baths, $A$ and $B$, have same temperature,
i.e. for $T_a = T_b = T$, $\psi(x)$ reduces to
\begin{eqnarray}
\psi(x) & = & \int^x \left \{ \frac{U^{\prime}(x) \Gamma(x)}{T
\gamma_a + T \gamma_b g ^2(x) + \gamma_b D_R \kappa_0^2 g^2(x)} \right. \nonumber \\
&& + \left. \frac {D_R \kappa_0^2}{\Gamma(x)} \left (\frac{
2\gamma_a\gamma_b g(x) g ^{\prime} (x)} {\gamma_a T + \gamma_b T g
^2(x)+ \gamma_b D_R \kappa_0^2 g^2(x) }\right ) \right \} dx. \nonumber \\
\end{eqnarray}\label{neq4}

\noindent Here also for linear coupling, i.e. for $f(x) = x$, there would be
no net current.

\section{General analysis}

Two interesting points may be noted here. For $U(x)=0$, i.e., even
in the absence of any external potential, fluctuation induced
directed motion is possible. The direction of current will be from
nonlinear bath to linear bath for $T_b=T_a=T$(say). Thus, the system
will act like a Carnot engine, which extracts work by making use of
two thermal baths not necessarily at different temperatures. Even
when both the baths are kept at $T = 0$, this engine operates due to
external modulation of the nonlinear bath which for
$\delta$-correlated external noise modulation, operates between two
temperatures $0$ and $D_0 \kappa_0^2$. On the other hand, in the
case of exponentially correlated external noise, the engine operates
between $0$ and $D_R$ where $D_R = D_e \kappa_0^2$. At this point it
is pertinent to mention that instead of mentioning the physical
temperature, the effective temperature induced by external driving
is the relevant measure. From the foregoing discussion it is evident
that the model shows the net motion occurs as long as the
\emph{relative effective temperature} difference between the two
baths ($A$ and $B$) is nonzero.

From the very mode of development it is clear that the magnitude of
the net current will depend on the slope of the effective potentials
$\phi(x)$ and $\psi(x)$. For $T = 0$ and $U(x) = 0$, the effective
slope of $\phi(x)$ is $(2f^{\prime\prime}(x)
\gamma_a)/(\Gamma(x)f'(x))$, whereas that for $\psi(x)$ is
\begin{eqnarray}
\psi(x) & = & \frac{2f^{\prime\prime}(x) \gamma_a}{\Gamma(x)
f^{\prime}(x)}-\frac{\tau_R \Gamma^{\prime}(x) \gamma_a}{\Gamma(x)
[1-\Gamma(x) \tau_R]} \nonumber \\
& \approx & \frac{2f^{\prime\prime}(x) \gamma_a}{\Gamma(x)
f^{\prime}(x)}- \frac{\gamma_a \tau_R
\Gamma^{\prime}(x)}{\Gamma(x)}.
\end{eqnarray}\label{neq5}

\noindent In the last step we have used the fact that for $\tau_R
\Gamma(x) < 1$.

When both $\Gamma(x)$ and $\Gamma^{\prime}(x)$ are positive (and
$\Gamma(x)\tau_R < 1$), the slope of the effective potential
$\psi(x)$ is less than that of $\phi(x)$ and consequently the net
flow reduces. Thus the magnitude of transport will be maximum if we
modulate the nonlinear bath by $\delta$-correlated noise. Any finite
correlation time will decrease the transport process for $\Gamma(x)
> 0$ and $\Gamma^{\prime}(x) > 0$. On the other hand for
$\Gamma^{\prime}(x)$ negative, the slope of $\psi(x)$ becomes larger
which increases the net flow. Thus for a given correlation time, the
magnitude of current will primarily depend on the relative sign of
$\Gamma(x)$ and $\Gamma^{\prime}(x)$. Also for a fixed value of
coupling function $f(x)$, the ratio of $\gamma_b$ and $\gamma_a$
determines the magnitude of the current. On the other hand when the
amplitude modulations of $f^{\prime}(x)$ and $f^{\prime\prime}(x)$
are small in comparison to the amplitude modulation of $U(x)$, the
slope of the effective potential $\phi$ will be approximately given
by $(U^{\prime}(x) \Gamma(x))/( \gamma_b D_0 \kappa_0^2 [f^{\prime}
(x)]^2)$ at $T =0$ and that of $\psi$ will be given by
$(U^{\prime}(x) \Gamma(x))/( \gamma_b D_R \kappa_0^2 g^2 (x))
\approx ((U^{\prime}(x) \Gamma(x))/(\gamma_b D_R \kappa [f^{\prime}
(x)]^2)) [ 1 + \Gamma(x) \tau_R ] $. This expression provides an
interesting fact that when the amplitude modulation of
$f^{\prime}(x)$ and $f^{\prime\prime}(x)$ are small, the net current
increases if we modulate the bath by exponentially correlated noise.
In this particular limit, the problem becomes equivalent to a
particle moving in a spatially varying temperature field, namely,
$T(x) = T \Gamma(x) + \gamma_b D_0 \kappa_0^2 [f^{\prime}(x)]^2$.
Almost three decades ago Landauer \cite{landauer} explored the
problem of characterizing non-equilibrium steady states in the
transition kinetics between the two locally stable bistable systems.
His idea was that the relative stability of a particle diffusing in
a bistable potential can be altered by an intervening hot layer
which has the effect of pumping particles from a globally stable
region to a metastable region. Latter this problem was carried in a
wider context by van Kampen and others \cite{vankampen2}. In absence
of any externally applied fields, B\"{u}ttiker \cite{buttiker}
suggested that the generation of current is an important consequence
of state dependent diffusion. In contrast, we address the problem of
Langevin equation with multiplicative noise and state dependent
diffusion for a thermodynamically open system to study the
non-equilibrium fluctuation induced transport phenomena.

\section{Conclusions}
In this article we have developed a microscopic model of a Maxwell
demon type engine. Our approach is based on the system-reservoir
model where the system is coupled with two baths. For one bath, the
system-reservoir coupling is bi-linear and for the other, the
coupling is non-linear in system coordinate which is externally
modulated by a noise agency. We then derive the Langevin equation
with a multiplicative noise and a nonlinear dissipation. Then by a
systematic expansion of the relevant variable in powers of inverse
of dissipation constant, we obtained the corresponding Smoluchowski
equation for state dependent diffusion in the limit of large
friction. We have applied the formalism to the problem of diffusion
of a particle in a periodic potential where the nonlinear bath act
as the source with an \textit{effective temperature}
and the linear bath as the sink and there by
providing a heat engine.

\begin{acknowledgments}
JRC and SC would like to acknowledge the UGC, Delhi
[PSW-103/06-07(ERO) and 32-304/2006(SR)] for financial support. SKB
acknowledges financial support from Department of Physics, Virginia
Tech.
\end{acknowledgments}

\end{document}